\documentclass[12pt]{article}
\usepackage{graphicx,amsmath,amssymb}
\usepackage{lineno}

\newlength{\dinwidth}
\newlength{\dinmargin}
\def\lapproxeq{\lower .7ex\hbox{$\;\stackrel{\textstyle                                                    
<}{\sim}\;$}}                                                    
\def\gapproxeq{\lower .7ex\hbox{$\;\stackrel{\textstyle                                                    
>}{\sim}\;$}}                                                    
\def\be{\begin{equation}}                                                    
\def\ee{\end{equation}}                                                    
\def\bea{\begin{eqnarray}}                                                    
\def\eea{\end{eqnarray}}
\begin{document}
\begin{flushright}                                                    
IPPP/20/2  \\                                                    
\today \\                                                    
\end{flushright} 
\vspace*{0.5cm}
\begin{center}
{\Large\bf Isolating the Odderon in central production  }\\
\vspace{.5cm}
{\Large\bf  in high energy $pA$ and $AA$ collisions}\\

\vspace{1cm}
R. McNulty$^{(a)}$, V.A. Khoze$^{(b,c)}$, A.D. Martin$^{(b)}$ and M.G. Ryskin$^{(b,c)}$ \\

\vspace{.5cm}
$^{(a)}$ School of Physics, University College Dublin, Dublin 4, Ireland.\\
$^{(b)}$ Institute for Particle Physics Phenomenology, University of Durham, Durham, DH1 3LE \\                                                   
$^{(c)}$Petersburg Nuclear Physics Institute, Kurchatov National
Research Centre,
Gatchina, St. Petersburg 188300, Russia\\

\end{center}

\vspace{0.5cm}
\begin{abstract}
\noindent
We study the rapidity dependence 
 of the central exclusive production cross sections of C-even  mesons in $pA$ and $AA$ collisions, where $A$ is a heavy ion. We observe qualitatively different behaviour of the contributions arising from $\gamma$-Odderon and Pomeron-Pomeron fusion mechanisms. This can be used to extract the Odderon signal from the events of $f_2$ mesons exclusively produced in the forward region.  Estimates, obtained using expected values of the Odderon cross section, indicate that the $\gamma$-Odderon contribution may exceed by a few times the Pomeron-induced background in Pb-Pb collisions. Moreover, the Odderon effect can be clearly seen in terms of the asymmetry in $pA$ and $AA$ collisions with the beam and target reversed. It is particularly interesting to note that the asymmetry for $\gamma$-Odderon fusion reaches its maximum value close to 1 in the forward direction, whereas the asymmetry for the Pomeron-Pomeron fusion contribution is small. The role of additional interactions of the $f_2$ meson with nucleons in the heavy ion, and also the contributions from secondary Reggeons, are estimated. The photon-Odderon contribution has a large normalisation uncertainty but the enhanced
cross-section in the forward region combined with a large asymmetry increases
 the chance
of experimentally detecting the Odderon.

\end{abstract}

\section{Introduction}
Central exclusive production (CEP) of C-even mesons was intensively discussed as a promising possibility for  searching for glueballs produced in Pomeron-Pomeron fusion ({see, for example, the reviews in \cite{Review}).  
Here we wish to discuss how
 this type of process can be used to search for the Odderon\footnote{The  Odderon is the odd-signature (C=-1) partner of the even-signature (C=1) Pomeron,
see for reviews \cite{Braun:1998fs,Ewerz:2003xi, Ewerz:2005rg}. It
is a firm prediction of QCD~\cite{Kwiecinski:1980wb,Bartels:1980pe}, but so far the experimental evidence for its existence is not definitive.
There have been various proposals on how to search for Odderon-exchange effects
in high energy collisions.  In particular,
there is a long history of studying
the possibility of searching for the Odderon via the exclusive photoproduction of C-even
mesons (such as the $f_2$);
see for
example \cite{Barakhovsky:1991ra}~-~\cite{Goncalves:2018pbr}, although no quantitative estimate of the background to the Odderon signal has been made.}.   In particular we study the possibility of searching for Odderon-exchange in ultraperipheral $pA$ collisions at the LHC, where $A$ is a heavy ion \cite{HKMR}. It was shown in \cite{HKMR} that the signal cross sections for the semi-exclusive production of C-even mesons due to Odderon-$\gamma$ fusion could be quite large, up to the $\mu$b level -- note that the heavy ion enhances the $\gamma$ flux by a factor $Z^2$. So, in principle, the observation of these could be viable search channels for Odderon exchange.  However it was also noted in \cite{HKMR} that as well as identifying a sizeable signal, it is essential to quantitatively estimate the contribution from all potential background processes. In addition to production from Odderon-$\gamma$ fusion we have irreducible backgrounds due to $\gamma\gamma$ and Pomeron-Pomeron fusion, and also a reducible background coming from the photoproduction of vector mesons followed by their radiative decay to the 
 C-even meson where the emitted photon is undetected.
If instead of $pA$ collisions, we were to study $AA$ collisions then the background from $\gamma\gamma$ fusion could be overwhelming, whereas in $pp$ collisions the background coming from Pomeron-Pomeron fusion would be strongly dominant. 
In ref.~\cite{HKMR} the signal and background were estimated for a set of C-even mesons
 $(\pi^0,~f_2(1270),~\eta,~\eta_c)$ produced exactly at the centre at rapidity $y=0$. In each case the background posed a serious challenge to the experiment. 
 Of these, the $f_2$ meson looked to be the most promising. 
The cross section is rather large  and the backgrounds due to $\gamma\gamma$-fusion and vector meson radiative decays (such as $J/\psi\to f_2~\gamma)$ are low\footnote{Note that in Table 3 of \cite{HKMR} the background caused by $\gamma\gamma$ fusion was underestimated by a factor of 5, but this contribution is still much lower than the Pomeron-Pomeron term.}. However the background due to Pomeron-Pomeron fusion still poses a problem.
 
In this paper, we examine these backgrounds further, as well as those arising from two other sources. In addition we consider the forward kinematics with $y=2-5$ which has some advantages in selecting the Odderon contribution, although here there are increased 
contributions from the exchange of secondary Reggeons. Since in the forward  (i.e large rapidity $y$) direction, the rapidity difference between the  $f_2$ meson and the nearest proton is not too large, the $R=\omega,~\rho$ trajectory exchange is not suppressed too much.

We also consider the viability of looking for an Odderon signal in heavy ion, $AA$, collisions.
The $\gamma\gamma$ fusion background can be controlled by requiring relatively large transverse momenta, $p_t>0.3-0.5$ GeV, of the observed $f_2$ meson 
 while the Pomeron-Pomeron fusion background in $AA$ case 
becomes less important.  We find that the relative size of the Odderon signal compared to the background is greater in $AA$ than in $pA$ collisions.

 We extend our studies in the search for the Odderon making use of the rapidity dependence of photoproduction reactions, to define asymmetries for exclusive $f_2$ production
in both $pA$ and $AA$ central processes.  The ability of the LHC to provide beams of protons and ions in either direction, means that forward detectors
like LHCb have an acceptance for $pA$ collisions at both positive and negative rapidities. 
We find a large asymmetry for $f_2$ production through photon-Odderon fusion that is largely absent in the background Pomeron-induced processes.

 We have to emphasize that in the $f_2\to\pi^+\pi^-$ decay mode there is a large background coming from the direct $\pi^+\pi^-$ ultraperipheral photoproduction and the higher mass tail of the $\rho$ meson which can be produced via the photon-Pomeron fusion with a quite large cross section. This background has no peak in the $f_2$ mass region but 
   experimentally it
   will strongly dilute the significance of the $f_2$ signal.
   
The outline of the paper is as follows. In Sections 2 and 3 we give the formulae for the cross sections for exclusive $f_2$ production in $pA$ collisions as a function of the $f_2$ rapidity arising respectively from Pomeron-Pomeron and Photon-induced fusion processes.
In Section 4 we note that the cross sections have an asymmetry in rapidity; the cross sections $\sigma(pA)$ and $\sigma (Ap)$, with interchanged proton and ion beams, are not equal in the forward direction.
In contrast the process $AA\to A+f_2+A$ is dominated by $\gamma\gamma$ fusion, and there is no asymmetry in rapidity. However, if we consider events where one of the ions $(A^*)$ is broken then 
an asymmetry is predicted to occur. Throughout the paper we therefore also present formulae for the process $AA\to A+f_2+A^*$.  Note that by $A^*$ we allow for the break up of the ion but not of the constituent nucleons inside the ion.
In Section 5 we consider the $\gamma\gamma$ background and emphasize that an important background to $f_2$ production in the $\pi^+\pi^-$ channel is from $\rho$ photoproduction.
In Section 6 we perform numerical calculations to give indications of the size of the cross sections and asymmetries expected for $pA\to p+f_2+A$ and $AA\to A+f_2+A^*$ by making physically reasonable assumptions for the unknown parameters. We consider both the Pomeron-Pomeron and the photon-induced fusion mechanisms.
The size of the Odderon induced cross section is unknown.  However,
by using $pA$ and $AA$ collisions and considering the asymmetry of the cross-section with
respect to rapidity, its effects can be enhanced.

\section{Rapidity dependence of Pomeron-Pomeron fusion}
Let us start with the pure exclusive $pp\to p+f_2+p$ reaction.  The cross section as a function of the rapidity of the $f_2$ meson has the form
\be\label{ac}
\frac{d\sigma^{\rm CEP}_{pp}}{dy}~=~\frac 1{16^2\pi^5}\int d^2p_1d^2p_2|A(p_1,p_2,y)|^2e^{-2(y_1-y_2)}
\ee
where $y_1\ (y_2)$ and $y$ are the rapidities of the beam (target) protons and $f_2$ meson respectively ($y_1>y>y_2$); and $p_1\ (p_2)$ are the transverse momenta of the outgoing protons; $t_1=-p^2_1,\ t_2=-p^2_2$.
The amplitude is dominated by double Pomeron exchange and reads
\be 
\label{a1}
A(p_1,p_2,y)~=~C\exp(Bt_1+Bt_2)~e^{\alpha_P(t_1)(y_1-y)}~e^{\alpha_P(t_2)(y-y_2)}\ ,
\ee
where $\alpha_P(t)$ is the Pomeron trajectory; $B$ accounts for the slope of the vertices
 and $C$ is the product of the coupling constants (two Pomeron-proton couplings times  the Pomeron-Pomeron-to-$f_2$ fusion constant).  For the Pomeron trajectory we use the simple form $\alpha_P(t)=1+\epsilon+\alpha'_Pt$
with $\alpha'_P=0.25$ GeV$^{-2}$ and $\epsilon=0.0808$ (corresponding to the Donnachie-Landshoff (DL) parametrization~\cite{DL}).
After integration over the transverse momenta in (\ref{ac}) the cross section becomes
\be\label{a2}
\frac{d\sigma^{\rm CEP}_{pp}(y)}{dy}~=~\frac{C^2}{16^2\pi^3}~~\frac{e^{2\epsilon(y_1-y_2)}}{4(B+\alpha'_P(y_1-y))(B+\alpha'_P(y-y_2))}\ .
\ee
The only rapidity dependence comes from the denominator of (\ref{a2}). Taking $B=8$ GeV$^{-2}$ (which is consistent with the WA102 data~\cite{WA102}) we expect a rather weak $y$-dependence. At $\sqrt s=13$ TeV the cross section increases by 1.3\% going from $y=2$ to $y=5$. This is negligible.

\subsection{$pA$ collisions}
A stronger $y$ dependence  is expected in heavy ion collisions when we account for the possibility of interaction of the $f_2$ meson with the ion (or rather with the nucleons, $N$, in the ion).  First we consider the exclusive process $pA\to p+f_2+A$. The probability of the above `secondary' interaction is driven by the $\sigma(f_2 N)$ cross section, which increases with energy, that is 
 with the $f_2$ rapidity as $\exp(2\epsilon(y-y_2))$  (here  the rapidity of target ion $y_2<0$ is negative).

Besides this we have to account for the additional beam-target inelastic interactions which populate the rapidity gap and thus violate the `exclusivity' condition. The corresponding `gap survival probability', $S^2$, for the case of collisions with a heavy ion was discussed in detail in ~\cite{SC3}. 
It is convenient to calculate the value of $S^2$ in impact parameter, $b_t$, space. For the case of survival against an additional proton-ion interaction~\footnote{We do not include the inelastic Glauber corrections since the effect of inelastic shadowing is almost compensated by the effect of short-range correlations in the wave function of the target nucleus~\cite{FSZ}.}
\be\label{s1}
S^2_{pN}(b_t)~=~\exp(-\sigma_{\rm tot}(pN)~T_A(b_t))\ ,
\ee  
where
$\sigma_{\rm tot}(pN)$ is the total cross section of the proton-Nucleon interaction and $T_A(b)$ is the optical density of the heavy ion,
\be\label{T}
T_A(b)=\int_{-\infty}^\infty dz (\rho_p(z,b)+\rho_n(z,b))\ .
\ee
In this simplified estimate we neglect the radius, $r_{pN}$, of the proton-nucleon interaction in comparison with the larger heavy ion ($A$) radius and consider the {\em total} exclusive cross section integrated over $t_1$ and $t_2$.

The nucleon density distribution in $A$ is described by the Woods-Saxon form~\cite{Woods}
\be
\rho_N(r)= \frac{\rho_0}{1+\exp{((r-R)/d)}}\;,
\ee
where the parameters  $d$ and $R$ respectively characterise the skin thickness and the radius of the nucleon density in the heavy ion; $r=(z,b)$. For $^{208}$Pb
we take the recent results of~\cite{Tarbert,Jones}
\begin{align}\nonumber
R_p &= 6.680\, {\rm fm}\;, &d_p &= 0.447 \, {\rm fm}\;,\\ \label{eq:pbpar}
R_n &= (6.67\pm 0.03)\, {\rm fm}\;, &d_n &= (0.55 \pm 0.01) \, {\rm fm}\;.
\end{align}
The nucleon densities, $\rho$, are normalized to 
\be
 \int\rho_p(r)d^3r=Z \;, \qquad \int\rho_n(r)d^3r=N_n\;,
\ee
for which the corresponding proton (neutron) densities are $\rho_0 = 0.063$ (0.093) ${\rm fm}^{-3}$.

Correspondingly the probability to preserve  the exclusivity against $f_2 N$ additional interactions with the target ion is given by
\be\label{s2}
S^2_{f_2N}(b)=\exp(-\sigma_{\rm tot}(f_2N)T_A(b))\ ,
\ee
where $\sigma_{\rm tot}(f_2 N)$ is the cross section of an $f_2$ meson interacting with a nucleon, $N$, in the ion.

 To calculate the rapidity dependence of $f_2$ production in proton-$A$ collisions we first compute the cross section for the incoherent process, $pA\to p+f_2 +A^*$, where the outgoing ion, $A^*$, breaks up. This cross section is given by that for the CEP in $pN$ collisions, $d\sigma^{\rm CEP}_{pp}/dy$, times the number of nucleons in the ion at fixed $b_t$ (i.e., $T_A(b))$ times the survival factors describing no additional $pA$ and $f_2 A$ interactions (i.e., the survival against the production of additional secondaries). Thus integrating over $b_t$ we have
\be\label{ucs2}
\frac{d\sigma^{\rm incoh}_{pA}}{dy}~=~\frac{d\sigma^{\rm CEP}_{pp}}{dy}\int d^2b 
 T_A(b)S^2_{f_2N}(b)S^2_{pN}(b)\ ,
\ee where 
$S^2_{pN}$ is given by (\ref{s1}) and $S^2_{f_2N}$ is given by ({\ref{s2}).
Strictly speaking we should account for the gap survival factor, $S^2_{pp}$, in  the proton-proton case as well. However in this case it can be included  in the constant 
$C$,  that is into the $d\sigma^{\rm CEP}_{pp}/dt$ value, which anyway should be taken from experiment.

The cross section for the coherent (`elastic') process, $pA\to p+f_2+A$ is a little more difficult to calculate. We have  first to calculate the amplitude, which is proportional to the total number of nucleons $(\int d^2b T_A (b))$ in the incoming ion. After integration over the 
 momentum $q_t$ transverse to the incoming ion we obtain the factor  $\delta (b-b^*)$, where $b$ and $b^*$ are the independent impact parameters of the contributing nucleons in the amplitude $A$ and the complex conjugate amplitude $A^*$, respectively.  Finally we obtain
 \be\label{cs2}
\frac{d\sigma^{\rm coh}_{pA}}{dy}~=~\frac{d\sigma^{\rm CEP}_{pp}}{dy}8\pi(B+\alpha'_P(y-y_2))\int d^2b 
 T^2_A(b)S^2_{f_2N}(b)S^2_{pN}(b),
\ee
where the optical density $T_A$ is now squared and the extra dimension due to the extra $T_A$ is compensated by
 the slope of the $NN$ cross section, $ 2(B+\alpha'_P(y-y_2))  $. The details of the calculation can be found in \cite{SC3}.

Note that the rapidity dependence is hidden in the energy/rapidity  behaviour of the slope $2\pi(B+\alpha'_P(y-y_2))$ and in the cross section 
$\sigma_{\rm tot}(f_2N)$ 
 that enters the survival factor (\ref{s2}).

\subsection{$AA'$ collisions}
For $f_2$ production in ion-ion collisions, $AA'\to A+f_2+A'$, we have 
 a third survival factor
\be\label{s3}
S^2_{f_2N'}(b_2)=\exp(-\sigma_{\rm tot}(f_2N')T'_{A'}(b_2))\ ,
\ee
where $N'$ denotes a nucleon in the target ion, while $T'_{A'}(b_1)$ is the target ion optical density.
As before, working in the $b_t$ representation, the central incoherent (with respect to  both $A$ and $A'$ ions) cross section for 
$AA'\to A^*+f_2+A^{'*}$ reads
\be\label{ucs1}
\frac{d\sigma^{\rm incoh}_{AA'}}{dy}~=~\frac{d\sigma^{\rm CEP}_{pp}}{dy}\int d^2b_1 d^2b_2
 T_A(b_1)T'_{A'}(b_2)S^2_{f_2N}(b_1)S^2_{f_2N'}(b_2)S^2_{NN'}(|\vec b_1-\vec b_2|)\ ,
\ee 
where the factor $S^2_{NN'}$ accounts for the rescattering  of constituent nucleons in the incoming beam and target ions.
(Recall that the incoherent cross section is normalised to pure central exclusive $pp$ collisions, which do not include proton excitations. Therefore in the above process $A^*$ means that the ion is broken but that the constituent nucleons in the ion remain intact.)
For ion-ion collisions the survival factor
\be\label{SAA}
S^2_{NN'}(b)~=~\exp(-\sigma_{\rm tot}(NN)\Omega_{AA'}(b))\ ,
\ee
with
\be\label{OAA}
\Omega_{AA'}(b)=\int d^2 b_1d^2b_2 T_A(b_1)T'_{A'}(b_2)\delta^{(2)}(\vec b-\vec b_1+\vec b_2)\ .
\ee
As shown in~\cite{SC3} (see Fig.4(right)) for the lead-lead interactions 
$S^2_{NN'}(b)\simeq\theta(b-17~{\rm fm})$; that is, it is close to the form of a $\theta$ function.

Correspondingly the coherent (with respect to $A$) $AA'\to A+f_2+A^{'*}$ cross section is obtained by a similar relacement in (\ref{cs2}) leading to the form
\be\label{cs1}
\frac{d\sigma^{\rm coh}_{AA'}}{dy}=\frac{d\sigma^{\rm CEP}_{pp}}{dy}8\pi(B+\alpha'_P(y_1-y))\int d^2b_1 d^2b_2
 T^2_A(b_1)T'_{A'}(b_2)S^2_{f_2N}(b_1)S^2_{f_2N'}(b_2)S^2_{NN'}(|\vec b_1-\vec b_2|).
\ee 
Recall that the rapidity dependence is hidden in the energy/rapidity  behaviour of the 
$\sigma_{\rm tot}(f_2N)$ cross section that enters the survival factors (\ref{s2},\ref{s3}).

We note that the values of the integral in (\ref{ucs2})  and (\ref{ucs1}) can be treated as effective numbers of nucleon-nucleon pairs (nucleons from the heavy ion) that produce the $f_2$ meson. The same is true for the factor $8\pi(B+\alpha'_P(y_1-y))$ times the integral (i.e. the whole r.h.s expression, except for the factor $d\sigma^{\rm CEP}_{pp}/dy$) in eqs. (\ref{cs2}) and (\ref{cs1})).
See~\cite{SC3} and sect.6.1 of~\cite{HKMR} for more discussion of the formulae in this section.

\subsection{Missing information in the survival factors}
Unfortunately the cross section $\sigma_{\rm tot}(f_2N)$ is not known. One possibility is to assume that it is equal to the pion-proton cross section described by the Donnachie-Landshoff parametrization~\cite{DL}
\be\label{s-pi}
\sigma_{\rm tot}(f_2 N)=\sigma_0(s/1~{\rm GeV}^2)^\epsilon
\ee
with $\sigma_0=13.6$ mb and $\epsilon=0.0808$. Another possibility is to say that $\sigma_{\rm tot}(f_2 N)=\sigma_{\rm tot}(\rho N)$ where the $\rho$-proton cross section is extracted from the $\rho$ meson diffractive photoproduction data~\cite{HERA} in the framework of the Vector Dominance Model (VDM)~\cite{VDM}. This gives $\epsilon=0.055$ and $\sigma_0=15.7$ mb, which defines the VDM form of $\sigma_{\rm tot}(f_2 N)$.
However even this value (which is a bit smaller in the relevant energy region) can be an overestimate. It is not excluded that the wave function of the $f_2$ meson  produced via Pomeron-Pomeron fusion has not at the outset its normal configuration, but rather is represented by the small size, $\hat r$, of the quark-antiquark pair which has a lower cross section ($\sigma\propto \alpha^2_s\langle\hat r^2\rangle$,~see \cite{BK}) than that of the finally formed meson in its `equilibrium' state. Therefore in our numerical estimates we will use also the absorptive cross section with $\sigma_0=15.7/2$ mb, half the value of that given for the $\rho$ meson by VDM.

\subsection{Including secondary Reggeon contributions}}
Besides $f_2$ production by  Pomeron-Pomeron fusion there are production amplitudes in which one or both Pomerons in the amplitude (\ref{a1}) can be replaced by a secondary Reggeon\footnote{Sometimes the secondary Reggeon, $R=f_2$, is said to lie on the so-called $P'$ trajectory.} of the form
\be 
\label{a1R}
A_R(p_1,p_2,y)=C_R\exp(B't_1+B''t_2)~e^{\alpha_R(t_1)(y_1-y)}~e^{\alpha_P(t_2)(y-y_2)}.
\ee
After the integration over the tranverse momenta (analogous to going from (\ref{a1}) to (\ref{a2})) we find that the rapidity dependence is now
is suppressed by the factor $\exp((\alpha_{f_2}-1)(y_1-y))$ or $\exp((\alpha_{f_2}-1)(y-y_2))$, due to a smaller intercept  $\alpha_R(0)=\alpha_{f_2}(0)\simeq 0.5$. Hence it is negligible for a large rapidity interval $y_1-y$. However in the forward direction, for example at a rapidity of 5 where the rapidity difference $y_1-y$ is 3.5-4.5, the interference of secondary Reggeon with Pomeron exchange may affect the rapidity distribution of the produced $f_2$ meson enlarging the cross section at larger $|y|$. The effect may be more important if for some reason the `Pomeron-Pomeron$\to f_2$' vertex is much smaller than the `Pomeron-$R\to f_2$' vertex.
Such a situation may occur if it happens that
the $f_2$ CEP follows a pattern of a purely
perturbative expectation, derived for the case
of the $2^{++}$ state formed by the heavy quarks, see
e.g. \cite{Khoze:2001xm,Kaidalov:2003fw}.
 
The contribution of secondary Reggeons should be clearly seen in the rapidity $y$-distribution as a fast growth of the cross section as $y$ increases towards the higher end of rapidity interval.
 
Note that within the  perturbative 
approach a heavy $2^{++}$ quarkonium
is produced by fusing gluons from the colliding Pomerons;
for a review see e.g. \cite{Harland-Lang:2014lxa}.
An important property of the perturbative CEP mechanism
\cite{Khoze:2000jm}
is that in the forward proton limit
the centrally produced state should obey 
the so-called $J_z^{PC}$=$0^{++}$ selection rule
($J_z$ is the projection of its spin onto the collision axis). 
If the zero helicity diphoton (digluon) transition to
the $2^{++}$ quark-antiquark state is suppressed
(which is
 true only in the non-relativistic 
quark approximation)
then we could expect the suppression of the tensor state CEP in the
proton-proton collisions.
Though a priori being far from obvious,
it was shown  (see e.g.\cite{Bergstrom:1982qv,Li:1990sx})
that even in the case of light quarks,
the helicity zero amplitude for the $\gamma\gamma$ coupling of
the $q\bar{q}$ tensor mesons
remains numerically small, and this was experimentally confirmed by the BELLE
collaboration \cite{Uehara:2008ep} in the high-statistics measurement
of the dipion production in photon-photon collisions.

 It is quite intriguing that while the $f_2(1270)$ CEP
was clearly seen in the ISR measurement at  $\sqrt s=62$ GeV
using the Split Field Magnet spectrometer \cite{Breakstone:1986xd,Breakstone:1988bm},
the $f_2$ signal disappears in the 
study with the Axial Field spectrometer at the same energy
but when the protons were scattered nearly forward
\cite{Akesson:1983jz,Akesson:1985rn}; that is here we are close to $J_z=0$ kinematics.
A vanishing of the $f_2$ signal at low momentum transfer to
scattered protons was observed also
in the E690 fixed target experiment at the Tevatron at
$\sqrt s=40$ GeV (see e.g. \cite{Gutierrez:2014yqa})
\footnote{We are grateful to Mike Albrow for bringing our attention
to this phenomenon,
see also \cite{Mike}.}.
A further indication along this line follows from the preliminary LHCb measurement \cite{LHCbR} of dipion central production in $p$Pb collisions at 8.16 TeV. While the $f_2$ signal is clearly seen when no special exclusivity requirement is imposed, it is strongly suppressed when there is no observed activity in the forward region.
Such a peculiar behaviour of the $f_2(1270)$ CEP at low momentum transfers
certainly needs further detailed investigation in particular
at the LHC energies with the dedicated forward proton detectors
TOTEM and ALFA.
Thus in comparison with the Pomeron-Pomeron CEP amplitude (\ref{a1}) the Pomeron-$R$ term most probably has a constant factor $C_R$ about a factor 1 to 4 larger than $C$.

\section{Rapidity dependence of photon-induced $f_2$ production}

C-even mesons can be produced in exclusive events either via
the fusion of two C-even objects (Pomeron-Pomeron) or two C-odd objects ($\gamma$-Odderon or $\gamma-R$ where $R=\rho$ or $\omega$) . The photon flux, $N_\gamma$, radiated by the lead ion is quite large - it is enhanced by a $Z^2=82^2$ factor and is a strong function of photon energy (rapidity). In $b_t$ space, which is convenient to account for the survival factors $S^2$,  the photon flux outside the heavy ion~\footnote{For small $b_t<R_A$ the contribution to the exclusive cross section is strongly suppressed by the gap survival factors $S^2$.} reads~\cite{flux}
\be
\frac{d^3N_\gamma}{dxd^2b_\gamma} ~=~ \frac{Z^2\alpha^{\rm QED}}{x\pi^2 b_\gamma^2}~ (xm_n b_\gamma)^2~K_1^2(xm_n b_\gamma).
\label{fluxb}
\ee
Here $K_1(z)$ is the modified Bessel function of the first kind; $x$ is the nucleon momentum fraction carried by the photon;  $b_\gamma$ is the $b_t$ position of the $f_2$ production vertex with respect to the centre of the ion; and $m_n$ is the nucleon mass. For large $
z$ the function $K_1(z)\propto e^{-z}$ decreases exponentially. However the values of $x$ relevant for central
$f_2$ production at the LHC are very small 
\be
x~\sim ~m_{f_2}~e^{-y}/\sqrt s~\sim 10^{-4}.
\ee
Note that  $z=1$ corresponds to $b_\gamma\gapproxeq 200$ fm. Thus the dominant contribution has a logarithmic 
$d^2b_\gamma/b^2_\gamma$ structure and comes from very large
$b_\gamma$ starting at $b_\gamma=R_A\ (2R_A$ for $AA$ collisions) and up to $b_\gamma\sim 1/xm_n$ for the case of proton-ion (ion-ion) collisions; $R_A$ is the ion radius.

\subsection{Photon-Odderon fusion}
Our particular interest is in $f_2$ production by $\gamma$-Odderon fusion so let us discuss this first.
For the large values of $b_\gamma$ mentioned above we can neglect the survival factors with respect to the ion which emits the photon. Hence we can put the survival factors
\be
 S^2_{NN'}=S^2_{pN}=S^2_{f_2 N}(b_\gamma)=1
 \ee
 and write the CEP cross section just as the product of the photon flux times the Odderon induced, $\sigma_{\rm Odd}$, cross section
\be\label{gs}
\frac{d\sigma_{\rm Odd}}{dy}~=~\frac{dN}{dy}~\sigma_{\rm Odd}(\gamma D\to f_2 D)\ ,
\ee
where $D$ denotes the proton in the case of $pA$ collisions or the ion $A'$ in the $AA'$ case.

Recall that the lead ion $A$ radiates the photon {\em coherently} and is not destroyed (otherwise we lose the large factor $Z=82$). On the other hand it is better to select the events of {\em incoherent} interactions with the ion $A'$.
In this way we suppress the background caused by $\gamma\gamma\to f_2$ fusion. The incoherent events can be selected by
observing the signal in the rapidity interval close to the $A'$ ion or by looking for the events with a relatively large transverse momentum of the $f_2$ meson, say, $p_{t,f_2}>0.3-0.4$ GeV. (Recall that $p_{t,\gamma}$ is still quite small due to the large values of $b_\gamma$; therefore $p_{t,f_2}$ is almost equal to the momentum transferred to $A'$.).

Since the Odderon intercept, $\alpha_{\rm Odd}$, is very close to 1
\cite{Braun:1998fs,Odd,Ewerz:2003xi}, the rapidity dependence of the photon-Odderon fusion cross section in $pA$ collisions is completely driven by the behaviour of the photon flux $dN/dy$. In the $AA'$ case 
 the gap survival factor $S^2_{f_2 N'}$ also has an effect. Indeed, the semi-exclusive $\gamma A'\to f_2 A^*$ cross section reads
 \be\label{fA}
 \sigma(\gamma A'\to f_2 A^*)~=~\sigma_{\rm Odd}(\gamma p\to f_2 p)\int d^2b T'_{A'}(b)S^2_{f_2 N'}(b)\ ,
 \ee
 where $A^*$ denotes the ion $A'$, after it was broken by the incoherent interaction, and $S^2_{f_2 N'}$ is given by (\ref{s3}).

\subsection{$\gamma $-$R$ fusion}
 Of course the Odderon exchange in (\ref{gs},\ref{fA}) can be replaced by C-odd secondary Reggeon $R=\omega$ 
 (or $R=\rho$) exchange. We get exactly the same expressions (eqs.(\ref{gs}) and (\ref{fA})). The only difference is that  the `elementary'  cross section $\sigma_R(\gamma p\to f_2 p)$ (which replaces $\sigma_{\rm Odd}$ in eqs.(\ref{gs},\ref{fA})) now depends on the $f_2$-proton energy, that is on the rapidity of the $f_2$ meson as
 \be\label{o-rho}
 \sigma_R(\gamma p\to f_2 p)~\propto ~\exp(2(\alpha_R-1)
 (y_1-y))\ .
 \ee
 This leads to a strong rapidity dependence of the secondary Reggeon-exchange contribution. At LHC energies this contribution is completely negligible at central rapidities
  (that is, $y$ close to 0 in the laboratory frame) but may reveal itself in the forward region where the difference $|y_1-y|$ becomes smaller and the exponential increase towards 1.
  
  Besides this there may be interference between different 
  contributions. The interference between the Pomeron and the Odderon is small since the Pomeron-exchange amplitude is mainly imaginary while the Odderon-exchange is real. On the other hand secondary Reggeon-exchange can interfere with both the Pomeron and the Odderon amplitudes. However below we will neglect the interference effects in our simplified numerical estimates.

\section{Asymmetry}
The differential cross-sections for exclusive $f_2$ production described above exhibit very different dependencies with rapidity.  That resulting from Pomeron-Pomeron fusion is rather flat, while photon-Odderon production has a strong dependence due to the photon flux. For proton-ion collisions this can be usefully encoded in an asymmetry, A, defined  as
\be\label{Ap}
{\rm A}(Ap)~=~\frac{\sigma(pA)-\sigma(Ap)}{\sigma(pA)+\sigma(Ap)}\ ,
\ee
where $\sigma(pA)$ and $\sigma(Ap)$ denote the cross sections
measured in runs with interchanged proton and ion beams (at the same $y_{f_2}$).

In $pp$ collisions and $AA$ collisions (where the ions remain intact) the asymmetry is absent (A=0). However
in ion-ion collisions we can have asymmetry by selecting events
where one ion ($A^*$) is broken while the other one ($A$) remains intact:
\be\label{Ai}
{\rm A}(AA^*)~=~\frac{\sigma(A^*A)-\sigma(AA^*)}{\sigma(A^*A)+\sigma(AA^*)}\ .
\ee

\subsection{Pomeron-Pomeron fusion}
Due to the small value of $\alpha'_P$, we may neglect the small 
 rapidity dependence of the $t$-slopes and hence the proton-nucleon amplitude (\ref{a1}) has no asymmetry. However, an asymmetry appears after we account for the survival factors $S^2$ in (\ref{ucs2}). Indeed, the probability to have no additional interactions of the $f_2$ meson with the nucleons inside the heavy ion decreases when the cross section $\sigma_{f_2 N}$ increases (see (\ref{s2})); i.e., when the rapidity difference $y-y_2$ (or $y_1-y$) becomes larger. This means that we expect a  larger cross section (\ref{ucs2}) in the case when the forward $f_2$ goes in the direction of the ion. 
 Let us denote this case as $(pA)$ so the corresponding asymmetry, A, is {\it positive}.

For ion-ion collisions the situation is a bit more complicated. We have the survival factors $S^2_{f_2 N}$ and $S^2_{f_2 N'}$ from both sides (both ions). However in (\ref{cs1}) the optical density $T_A(b_1)$ is squared. That is
the typical values of $T_A(b_1)$ in the unbroken ion are larger than, $T_{A'}(b_2)$, in the ion that was destroyed. Therefore the factor $S^2_{f_2 N}(b_1)$ becomes more important and the $f_2$ meson  would prefer to fly in the direction of the unbroken ion (which interacts coherently).

\subsection{Photon-Odderon fusion}
For photon induced processes the cross section is proportional to the photon flux $N_\gamma$ (\ref{fluxb}), which increases with $x$ decreasing. This effect is stronger than that caused by the $S^2$ factors. Therefore now the cross section is larger when the $f_2$ meson goes in the direction opposite to the ion that was not destroyed and `coherently' radiates the photon. This leads to a {\it negative} asymmetry A.

\subsection{Fusion with a secondary Reggeon}
Recall that for proton-ion collisions at the LHC the energy per nucleon for the lead beam is about 2.5 times smaller than the proton beam energy. This leads to an additional asymmetry. Since the $R$-exchange cross section decreases with energy (that is with the rapidity difference) the $R$-exchange  contribution is larger for kinematics in which the $f_2$ meson goes in the ion direction; hence adding some {\it positive} component to the asymmetry A$(pA)$.

\section{Backgrounds}
When searching for the Odderon contribution in exclusive $f_2$ production we face two obvious sources of background. These arise from the production of the $f_2$ meson by Pomeron-Pomeron fusion and $\gamma\gamma$ fusion. In the next Section we will give indicative estimates of the size of the contribution arising from Pomeron-Pomeron fusion using the formalism that we developed in Section 2.  We first discuss the background from $\gamma\gamma$ and $\gamma$-Pomeron fusion.

\subsection{$\gamma \gamma$ fusion}
 It must be mentioned that Odderon exchange can be replaced by the photon exchange. Such a photon-photon fusion contribution can be calculated with rather good accuracy based on the known $f_2\to \gamma\gamma$ decay width. This gives
 \be
 \label{gamgam}
 \frac{d\sigma_{\rm QED}(\gamma p\to f_2+p^*)}{dt}~=~\frac{0.23\mbox{nb}}{|t|}F^2_{\gamma\gamma\to f_2}(t)\ ,
 \ee
 where $p^*$ indicates that we allow the proton to dissociate
 into some low mass state $p^*$ (since on this side we are looking for the `incoherent' process). For this reason we omit the proton form factor in (\ref{gamgam}).
 
 Practically it is impossible to distinguish here between the photon and the Odderon exchanges. Formally in the case of a photon we have an extra $1/t$ factor in (\ref{gamgam}) and may expect a steeper $t$-dependence. On the other hand we do not know the   $F^2_{\gamma\gamma\to f_2}(t)$ form factor and already we have selected not too small $|t|$. Note that 
 the trivial $\gamma\gamma$ contribution
 will have the same $y$-behaviour as that for Odderon exchange.  Integrating over the $0.3<p_{t,f_2}<1$ GeV, it must be normalized to $\sigma\simeq 0.3-0.4$ nb instead of
 $\sigma_{\rm Odd}(\gamma p\to f_2 p) =1$~nb which will be used for the numerical estimates in Section 6. The value taken for the cross section $\sigma_{\rm Odd}$ is discussed in Section 6.1.

\subsection{$\gamma$-Pomeron fusion}
   For photon-induced production, the $f_2$ peak  
   is placed on the top of a large background coming from the tail of $\rho$(770) meson ultraperipheral photoproduction (the $\rho$ meson is produced via the photon-Pomeron fusion with a rather large cross section). In particular, taking $\sigma(\gamma p\to \rho p)\simeq 10~\mu$b measured at HERA \cite{wing}, we expect the $\rho$-photoproduction induced $\pi^+\pi^-$ cross section, in the interval $M(f_2)\pm \Gamma(f_2)/2$, to be $\sigma \simeq 270$ nb.  
How does this compare with $f_2\to \pi^+\pi^-$ production via Odderon exchange? There are, at present, no data for $\sigma (\gamma p\to f_2 p)$, but experimental limits of 16 nb~\cite{HERAf2} indicate that the value is much smaller than from the tail of the $\rho$ resonance. We discuss this further in Section 6.1.

  Clearly, in order to separate out a pure $f_2$ signal it would be desirable to perform the partial wave analysis selecting a $J^P=2^+$ state.  However with such a large contribution from other partial waves it would be difficult and require very large statistics.
  A possible way to avoid the serious $\rho$ background would be to seek events for the $f_2$ meson via its $\pi^0\pi^0$ and $KK$ decay modes.

\section{Numerical estimates}
To get an impression for the size of the cross sections and asymmetries we present in Figs.~\ref{f1}-\ref{f3} the results of some numerical  calculations. These plots are shown for illustration. Unfortunately, as  mentioned, we do not know the size of the Odderon-photoproduced cross section, $\sigma_{\rm Odd}$; nor the values of the required couplings, like 
\be
{\rm PP}\to f_2,~~~{\rm P}R\to f_2,~~~\gamma{\rm O}\to f_2,~~~\gamma R \to f_2,
\ee
where P, O and $R$ denote Pomeron, Odderon and Reggeon, respectively.  Also we do not know the cross section $\sigma_{\rm tot} (f_2 N)$ where $N$ is a nucleon.
 Therefore we plot the contributions of the different components separately. We consider `13 TeV' kinematics, that is the proton beam has 6.5 TeV energy while the energy of the nucleon in the lead ion is 2.56 TeV.

 \subsection{Input assumptions and notation for the curves in the figures}
 
The cross section for the photoproduction of the $f_2$ meson by Odderon-exchange is the largest unknown in our predictions concerning the proposed search for the Odderon in $pA$ and $AA$ collisions.  A reasonable rough estimate is
\be
\sigma_{\rm Odd} (\gamma p\to f_2p^*)~~\sim~~ 1- 10~{\rm nb}.
\ee  
  Expectations based on lowest-order QCD give values in the region of 1 nb, whereas HERA data \cite{HERAf2} give an upper limit of 16 nb. For our numerical estimates below we will normalize our predictions to 1 nb; these are then easy to scale up or down as appropriate.
  
 To evaluate the cross section of central $f_2$ production via Pomeron-Pomeron fusion  we normalize the first factor, $C$, in (\ref{a1}) to be in agreement with the CMS data~\cite{CMS}. Thus we take 
 \be
 d\sigma (f_2)/ dy=1~\mu{\rm b ~~~at~~~} y=0.
 \ee In such a form it will be easy to recalculate the result expected in the case of another value of $C$ or $d\sigma(f_2)/dy$.
 For the $R$-contribution we put $C_R=2C$ and $\alpha_R(0)=1/2$ in (\ref{a1R}).
  Thus the ratio of the fusion amplitudes is given by
  \be
  \label{RP}
  \frac{{\rm P}R~{\rm amplitude}}{\rm PP ~amplitude}~~=~~2/\sqrt{s_{f_2 N}/1\mbox{GeV}^2}.
  \ee
  
   The absorptive cross section $\sigma_{\rm tot}(f_2N)$ is chosen in three different ways: via the VDM approach (\ref{s-pi}) with $\sigma_0=15.7$ mb, with  a twice smaller $\sigma_0=15.7/2$ mb (taking $\epsilon=0.055$), and with no absorption inside the heavy ion ($\sigma_0=0$).
  
   The $\gamma$-Odderon induced cross section is normalized to 
 $\sigma(\gamma+p\to f_2+p^*)=1$ nb.
 We use the VDM to evaluate the Reggeon contribution. It gives
 \be
 \label{RO}
 \frac{\gamma R~{\rm amplitude}}{\gamma {\rm O ~amplitude}}~~\simeq ~~5/\sqrt{s_{f_2 N}/1\mbox{GeV}^2},
 \ee
 where here the $R$-Reggeon is $\omega,\ \rho$. 
 
To search for the Odderon we have to keep one ion unbroken in order to have coherent photon radiation. Therefore we consider $AA^*$ and/or $Ap$ final state configurations.
  However the $A^*A^*$ contribution (with both ions incoherent) is also shown in Fig.~\ref{f1} by the dashed curve for comparison.

   Note that, to enlarge the statistics, in our estimates we allow also some low mass excitations of the nucleons. That is, when discussing CEP processes we bear in mind CEP$^*$ (which includes low mass excitations similar to \cite{HKMR}).
In all the figures the rapidity of the $f_2$ meson is defined to be positive if the $f_2$ is going in the direction of the proton for $pA$ collisions and in the direction of $A^*$ for the $AA^*$ case.

  \begin{figure} [h]
  \hspace{3.3cm}
\includegraphics[scale=0.5]{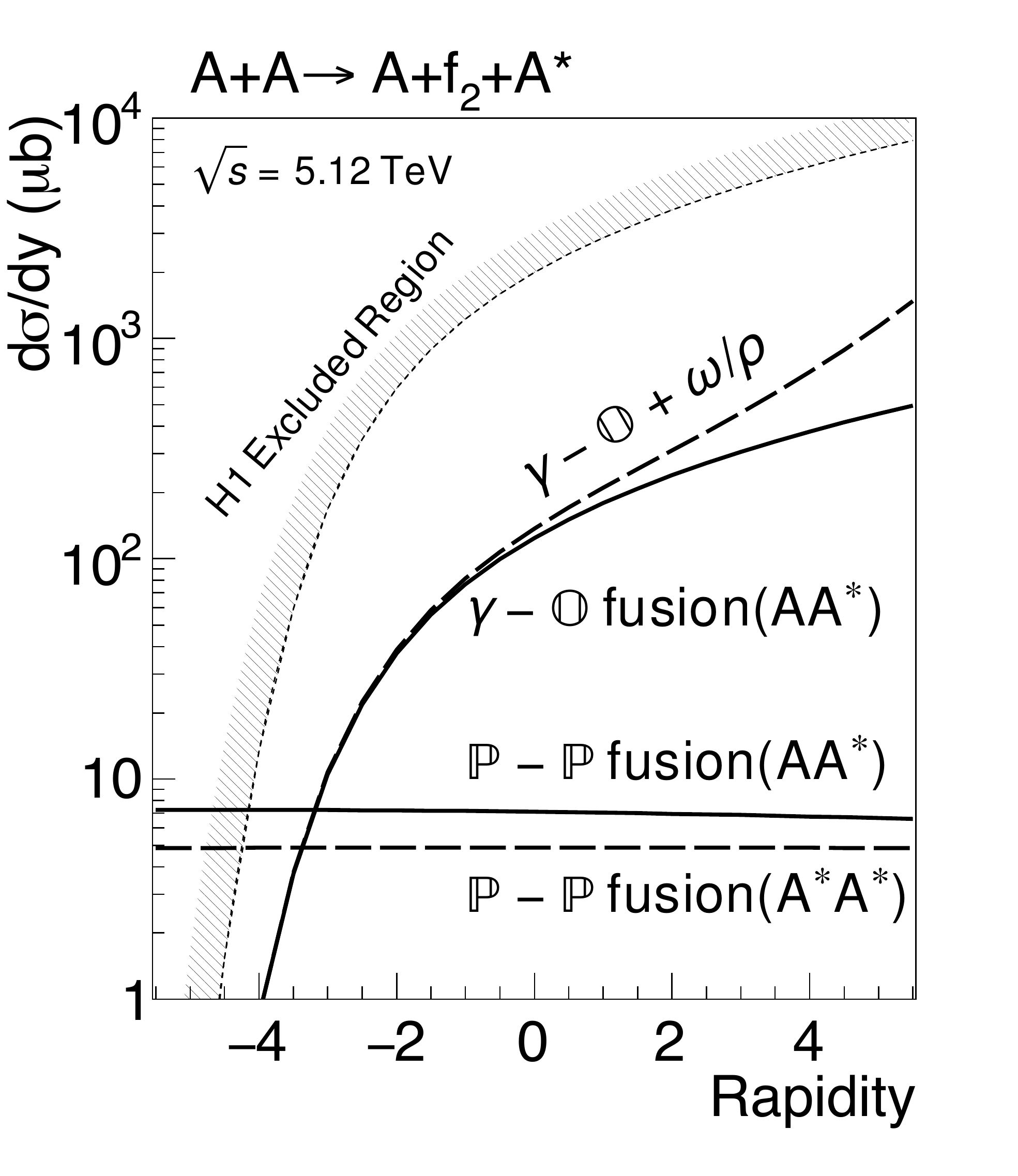}
\caption{\sf Indicative predictions for the photon- and Pomeron-induced cross sections for the (incoherent) process $AA\to A+f_2+A^*$ as a function of the $f_2$ rapidity for $\sqrt{s_{NN}} =5.12$ GeV. In this plot the $f_2$ absorptive cross section $\sigma_{\rm tot}(f_2 N)$ in the heavy ion is calculated from (\ref{s-pi}) with $\sigma_0=15.7/2$ mb and $\epsilon=0.055$. The dashed curve for $(A^*A^*)$ is  shown only because this process serves as a possible background to the Pomeron-induced $(AA^*)$ contribution. The shaded band indicates the region predicted to be excluded for the Odderon signal if we were to use the upper limit for $\sigma (\gamma p \to f_2 p)$ of 16 nb found at HERA\cite{HERAf2}, rather than 1 nb.}
\label{f1}
\end{figure}

\vspace{-2cm}
\begin{figure} [h]
  \hspace{-.5cm}
\vspace{-1cm}
\includegraphics[scale=0.45]{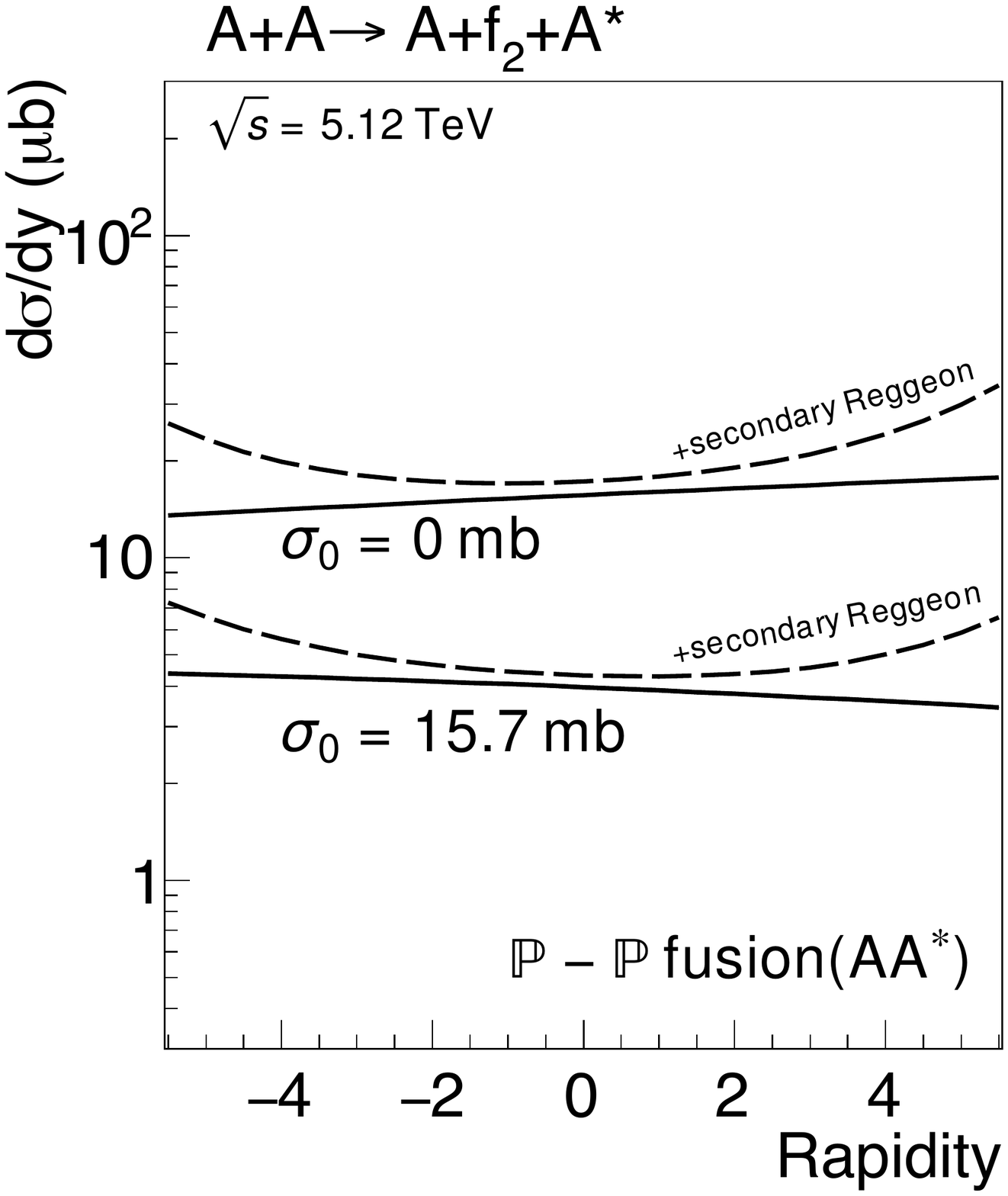}
 \includegraphics[scale=0.45]{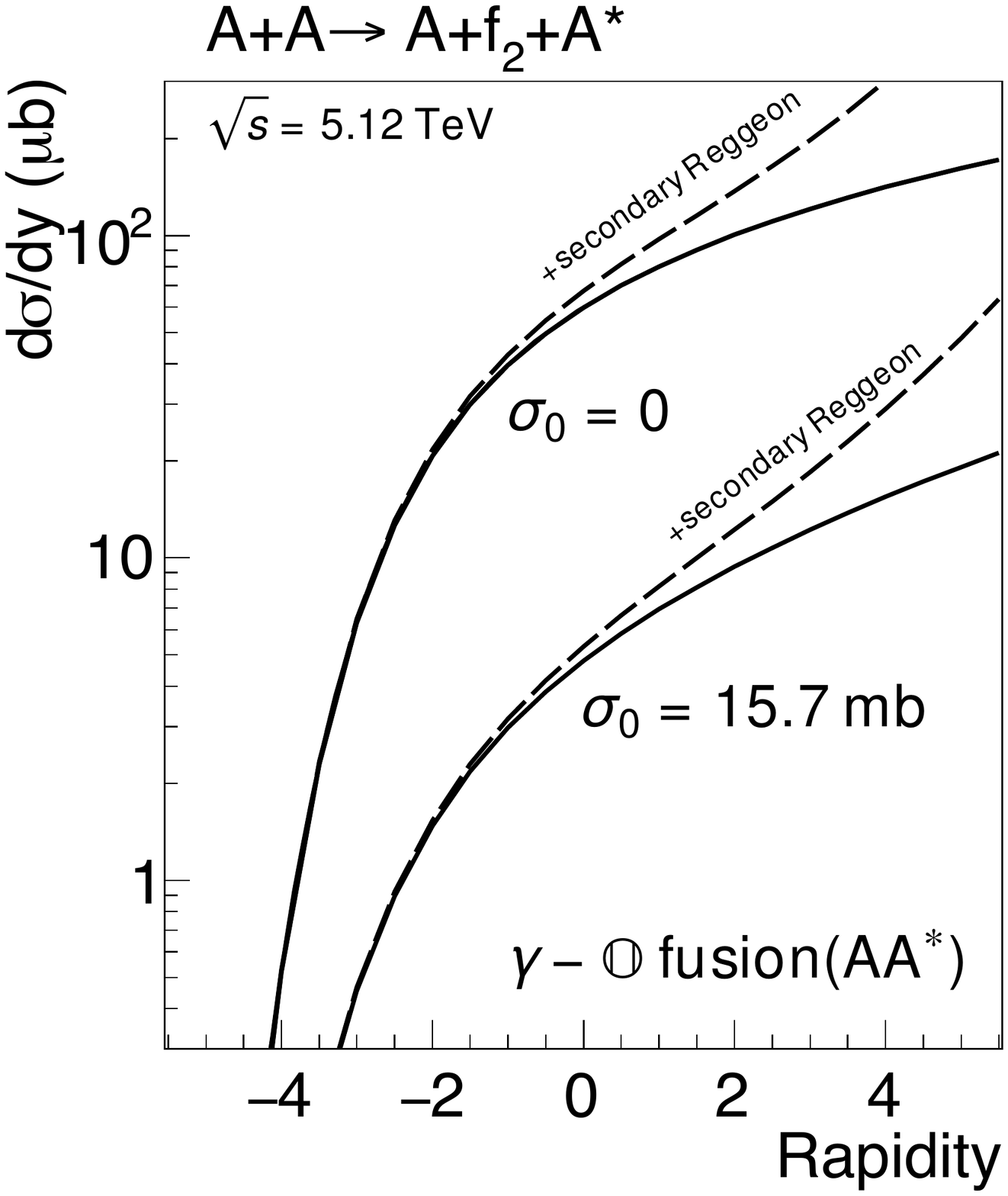}
\caption{\sf The shift of the continuous curves show the effect of changing $\sigma_{\rm tot}(f_2 N)$ from 0 to the VDM form of (\ref{s-pi}) with $\sigma_0$=15.7 mb and $\epsilon=0.055$. The dashed curves show the possible effect of including a secondary Reggeon contribution. Note that the two plots, showing respectively the Pomeron- and photon-induced contributions to $AA\to A+f_2+A^*$, have the same scale.} 
\label{f2}
\end{figure}

\begin{figure} [h]
  \hspace{2.3cm}
 \includegraphics[scale=0.5]{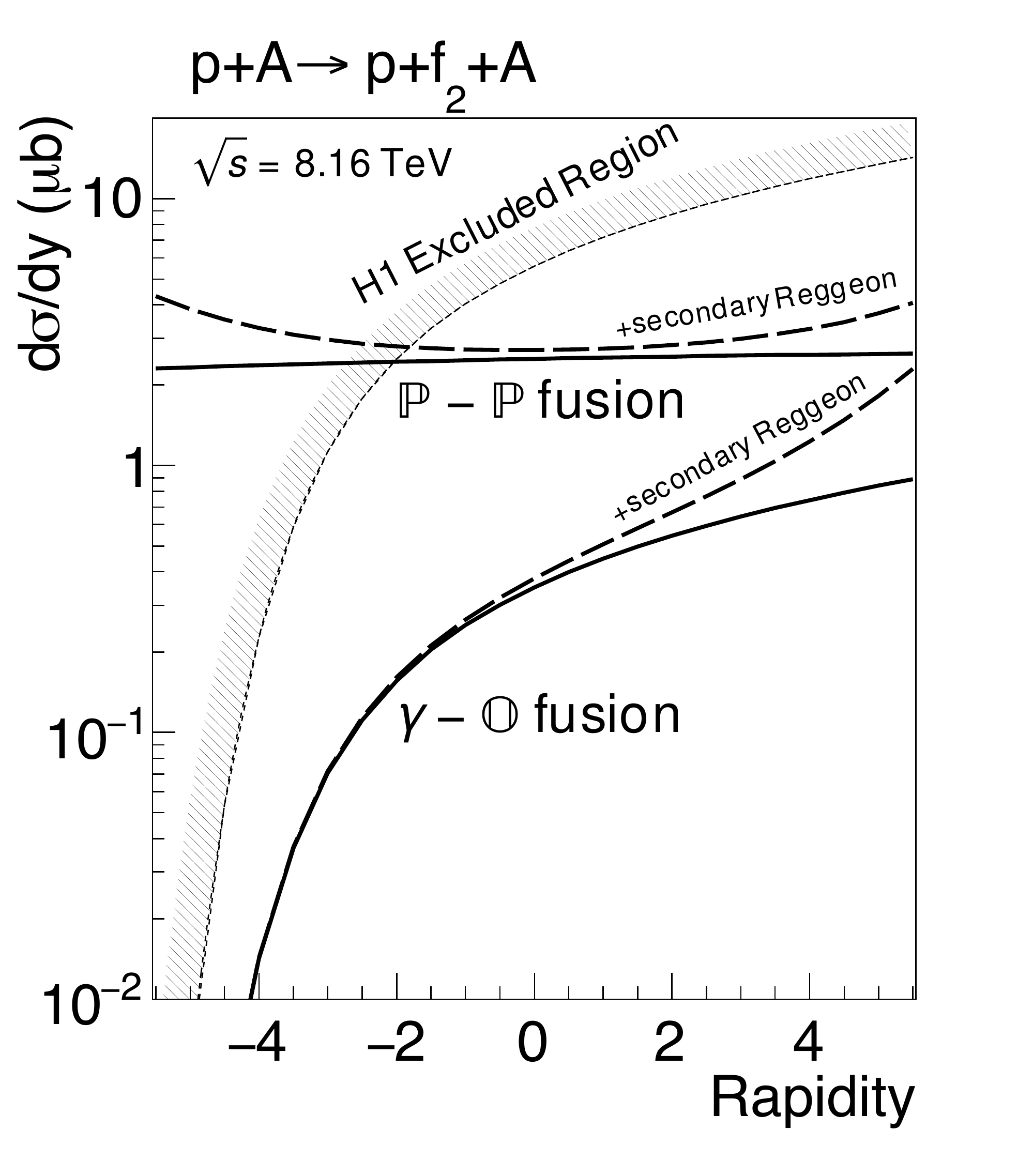}
\caption{\sf Indicative predictions of the Pomeron-Pomeron fusion and $\gamma$-Odderon fusion contributions to the cross section for $pA\to p+f_2+A$ as a function of the $f_2$ rapidity for $\sqrt{s_{NN}} =8.16$ GeV. The dashed curves show the effect of including a secondary Reggeon contribution. The effect of changing the absorptive cross section $\sigma_{\rm tot}(f_2 N)$ has a smaller effect than that for $AA'$ collisions which was shown in Fig.~\ref{f2}. Here we take $\sigma_{\rm tot}(f_2 N)$ from (\ref{s-pi}) with $\sigma_0$=15.7/2 mb and $\epsilon=0.055$. The shaded band indicates the region predicted to be excluded for the Odderon signal if we were to use the upper limit for $\sigma (\gamma p \to f_2 p)$ of 16 nb found at HERA\cite{HERAf2}, rather than 1 nb.}
\label{f4}
\end{figure}

\begin{figure} [h]
 \vspace{-2cm}
 \includegraphics[scale=0.43]{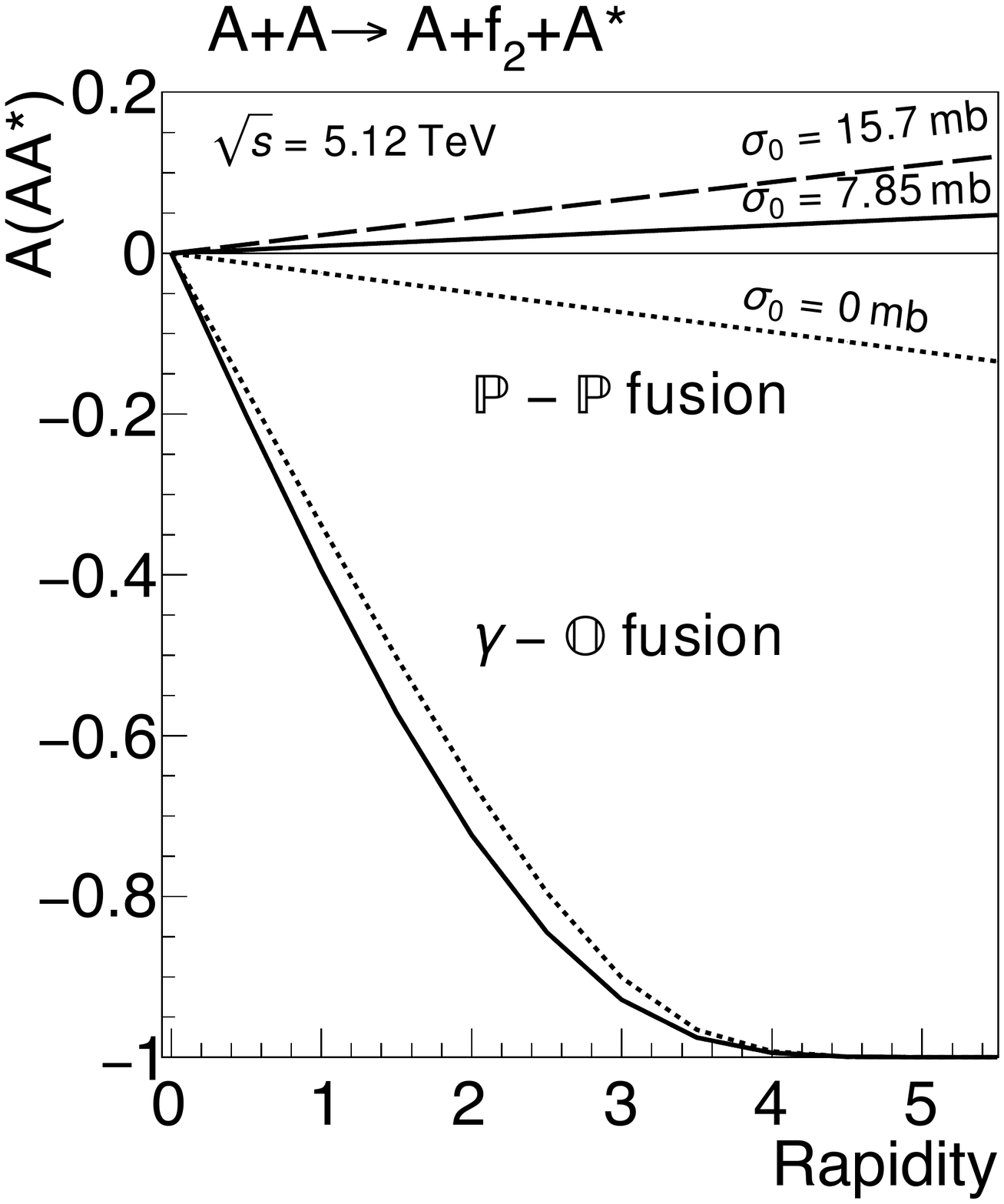}
 \includegraphics[scale=0.43]{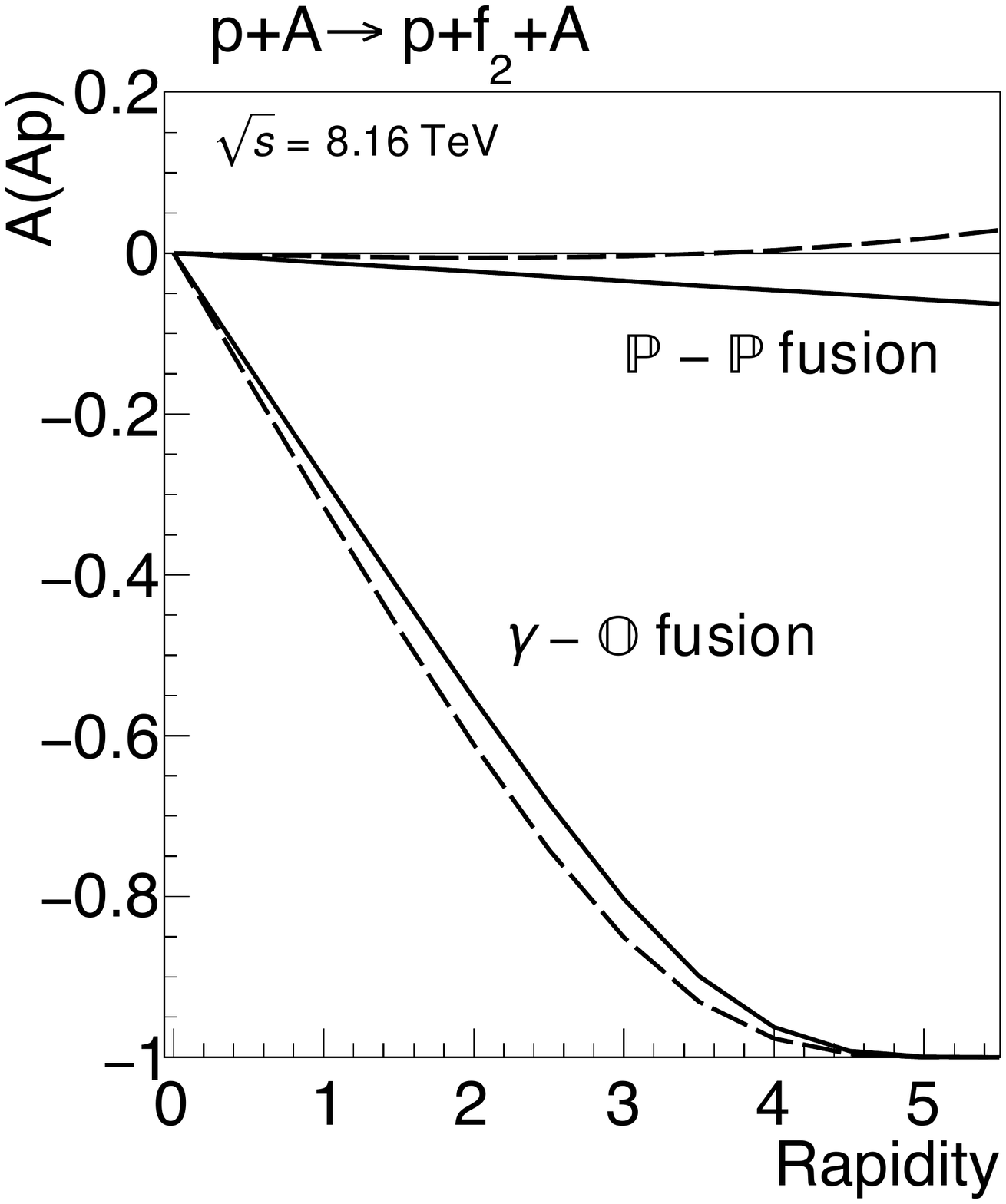}
 \vspace{-2cm}
\caption{\sf Predictions for the Pomeron-Pomeron and $\gamma$-Odderon fusion contributions to  the asymmetry as a function of the rapidity of the $f_2$ meson produced in the process $AA\to A+f_2+A^*$ (that is (\ref{Ai}) plotted in the left-hand diagram) and in the process $pA\to p+f_2+A$ (that is (\ref{Ap}) plotted in the right-hand diagram). We also show the effect of changing $\sigma_{\rm tot}(f_2 N)$ from 0 to forms given by (\ref{s-pi}) with $\sigma_0$=15.7/2 or 15.7 mb and $\epsilon=0.055$.  The effects of including secondary Reggeon-exchange terms only change the predictions within the limits of the $\sigma_0$ predictions.}
\label{f3}
\end{figure}

\subsection{Results for the cross section and the asymmetry}
The figures are based on the parameters described above.  
Figs.~\ref{f1} and \ref{f2} show the contributions to the cross section $d\sigma/dy$ for the process  $AA\to A+f_2+A^*$, while Fig.~\ref{f4} shows the results for $pA\to p^*+f_2+A$.  The contributions to the asymmetries in $f_2$ rapidity for the respective processes are compared in the two plots of Fig.~\ref{f3}. 

We start with the discussion of Fig.~\ref{f1}. It shows the exclusive production of the $f_2$ ($y=-5.5 $~to~$ 5.5$) in PbPb collisions. Assuming (\ref{gs}) with $\sigma_{\rm Odd}=1$ nb, the Odderon-induced cross section\footnote{For comparison we also show by the shaded band the upper limit of the Odderon-exchange signal if we were to use the HERA H1 limit of 16 nb \cite{HERAf2} for the  $f_2$ photoproduction cross section via  Odderon exchange.} in the forward region 
 is an order-of-magnitude larger  than that driven by Pomeron-Pomeron fusion. Secondary Reggeons (shown by dot-dashed curve for $\gamma$-induced production) may noticeably enlarge (about 2 times at $y=4$) the cross section in the forward region, where the suppression of the amplitude, exp$((y_1-y)/2)$ is not too strong, while the coupling to the secondary Reggeon may be quite large, see (\ref{RO}). 

The dependence of the predictions on the $f_2$ absorption cross section is shown in Fig.~\ref{f2}.
Of course larger absorption leads to a smaller cross section of $f_2$ meson production, as seen from the figure.
Note that for the $\gamma$-induced component we have stronger absorption. Indeed, for Pomeron-Pomeron fusion the major contribution comes from the region $b_1,~b_2\sim R_A$, see (\ref{cs1}).  That is the $f_2$ is created on the periphery of both ions where the optical density $T_A(b)$ is not large. On the contrary for the $\gamma$-induced component we deal with large $b_\gamma$; that is, there is practically no absorption by the ion $A$, but the integral over the parameter $b_2$ ($b$ in (\ref{fA})) covers the whole transverse area of the ion $A^*$. Hence in the survival factor (\ref{s3}) we have a much larger optical density $T_{A^*}$. Therefore the dependence of the $\gamma$-induced curves on $\sigma_0$ is greater.

The dashed curves in Fig.~\ref{f2} show the possible effect of adding the secondary Reggeon amplitude. Again the effect is stronger in the $\gamma$-induced case since we expect a larger $\gamma Rf_2$ coupling for the  fusion with a secondary Reggeon, see (\ref{RP}) and (\ref{RO}). 

Fig.~\ref{f2} clearly illustrates the importance of the survival factors with respect to the $f_2$ additional interactions with the nucleons in the heavy ion, both on the value of the cross section and its rapidity dependence. Note also that for $\sigma_{\rm tot}(f_2N)=0$ in Fig.~\ref{f2}(a), that is when $S_{f_2N}=1$ the cross section increases with rapidity, while it decreases if we take the VDM form given by (\ref{s-pi}) with $\sigma_0$=15.7 mb and $\epsilon=0.055$.

The contributions to the cross section for the process $pA\to p+f_2+A$ are plotted in Fig.~\ref{f4}. Here we take $\sigma_0=15.7/2$ mb and $\epsilon$=0.055 in (\ref{s-pi}), and take the $pA$ collisions to have an energy $\sqrt {s_{NN}}$=8.16 GeV. For this process we have no enhancement of the $\gamma$-induced contributions, that previously was observed in Fig.~\ref{f2}, due to the large transverse surface of the target $A^*$. Therefore the Odderon contribution is expected to be a few times smaller than that due to Pomeron-Pomeron fusion.

Fig.~\ref{f3} shows that the asymmetry of the Pomeron-induced cross section is small. Without secondary Reggeons it is caused by the non-zero slope, $\alpha'_P$ of the Pomeron trajectory (and gives a small negative asymmetry A$(AA^*)<0$) while the energy dependence of the absorptive cross section gives a small positive asymmetry (A$(AA^*)>0$). Recall that without the $f_2$ absorption the secondary Reggeon contributions do not produce an additional asymmetry. 
For the photon-induced component we observe in the forward region  ($y>0$) a large negative asymmetry that already by $y=3$ is close to -1 due to the growth of the photon flux (\ref{fluxb}) as $x\propto e^{-y}$ decreases.

For $pA$ collisions, the same qualitative behaviour of the asymmetry is observed as in $AA$ collisions, although the asymmetry for the Pomeron-induced process is slightly smaller, and the photon-induced asymmetry approaches -1 at slightly larger rapidities, as seen by comparing the plots in Fig.~\ref{f3}.

Recall, however, the possibility mentioned in Section 2.4 that the Pomeron-Pomeron fusion background may be suppressed. The dominant Pomeron-induced background will then be caused by fusion with a secondary Reggeon. In such a case there will be a large positive asymmetry in Fig.~\ref{f3} (right) since when the $p$ beam energy is larger than the energy of the nucleon in the ion, the dominant diagram is that where the secondary Regeeon couples to the ion $A$ and thus the $f_2$ meson goes in the direction of ion.

In summary, assuming an Odderon cross section $\sigma_{\rm Odd}=1$ nb in (\ref{gs}), the exclusive $f_2$ photo-production cross-section in the forward region for 
 $AA\to A+f_2+A^*$ processes 
 is expected to be an order-of-magnitude 
 larger than for Pomeron-Pomeron fusion in $AA$ collisions already at $y=2$ where the secondary Reggeon(s) contribution is still small. 
 Now the major background is caused by the trivial $\gamma\gamma\to f_2$ process of subsection 5.1. 
 Thus in the $AA$ case we have a chance to observe the Odderon signal in the $f_2\to \pi^0\pi^0$ mode if the corresponding $\sigma_{\rm Odd}$ cross section exceeds 0.3-0.5 nb~\footnote{Note that the $\gamma\gamma\to f_2$ contribution can be well controlled. The $f_2\to\gamma\gamma$ width is well known while the photon flux for the particular kinematics of the experiment can be monitored via the pure QED $\mu^+\mu^-$ pair production process.}.
  
 In $pA$ collisions the expected Odderon signal is a factor of 5 smaller than that due to Pomeron-Pomeron fusion.  The asymmetries A(${pA})$ and A$({AA^*})$ are predicted to be similar in all photo-produced processes, and are largely absent in Pomeron-Pomeron fusion.

 \section{Conclusions}
 The dependence of the cross section on the rapidity of a centrally produced meson is studied in proton - heavy ion $(pA)$ and in heavy ion-ion $(AA)$ interactions in (semi) exclusive processes (CEP$^*$). We consider the contributions due to production by Pomeron-Pomeron, Pomeron-Reggeon, $\gamma$-Odderon, $\gamma\rho$ and $\gamma \omega$ fusion.
The presence of a secondary Reggeon is found to be noticeable in the forward direction of the produced meson, especially when the meson rapidity becomes close to that of the heavy ion. Recall that with LHC kinematics the rapidity of a heavy ion beam is about one unit smaller than that for the proton beam. Thus in the forward direction the rapidity  difference between the nucleon $N$ in the ion and the produced meson is not large and the contribution of the secondary Reggeon is not sufficiently suppressed.

We emphasize that the additional interactions of the produced meson with the nucleons in the ion fills the rapidity gap and destroys the exclusivity of the events. This effect is encoded in the gap survival factors. Since the cross section of an additional interaction depends on the meson-nucleon energy, this leads to an additional rapidity dependence of the CEP$^*$ cross section. However, all these effects are much weaker than the rapidity dependence of the photon flux emitted by the heavy ion.  As an example we present estimates of the cross section for C-even $f_2$ meson. The CEP of a C-even $f_2$ meson can arise from the fusion of either two C-odd exchanges or two C-even exchanges. That is respectively $\gamma$-Odderon or Pomeron-Pomeron fusion. The difference in the rapidity  behaviour of these two contributions to $f_2$ production helps to extract the Odderon signal. 

Of course, the Odderon couplings are not known. However reasonably justified values (by matching with lowest-order QCD calculation \cite{HKMR}) provide estimates which show that, in the process $AA\to A+f_2+A^*$, the Odderon-induced signal may exceed by a few times the Pomeron-induced background.  On the other hand for $pA$ collisions, $pA\to  p+f_2+A$, the expected Odderon signal is a few times smaller than that due to the Pomeron-Pomeron background.
 
 A particularly interesting possibility to reveal the difference between the two production mechanisms is to measure the forward-backward asymmetries, (\ref{Ap}) or (\ref{Ai}), corresponding to the interchange of the proton and the ion or of the broken ($A^*$) and unbroken ($A$) ions. We see from Fig.~\ref{f3} that for $\gamma$-Odderon fusion the asymmetry approaches its maximum value in the forward direction, unlike the behaviour for Pomeron-Pomeron fusion where the asymmetry is less than about 0.1.  

\section*{Acknowledgements}
We thank Tara Shears for valuable discussions.
 RMcN and MGR thank the IPPP at the University of Durham for hospitality.

  A.~Kirk,
  Int.\ J.\ Mod.\ Phys.\ A {\bf 29}, no. 28, 1446001 (2014)
  [arXiv:1408.1196 [hep-ex]];\\

  M.~Albrow,
  Int.\ J.\ Mod.\ Phys.\ A {\bf 29}, no. 28, 1446014 (2014);\\
 
  M.~Albrow,
  AIP Conf.\ Proc.\  {\bf 1819}, no. 1, 040008 (2017)
  [arXiv:1701.09092 [hep-ex]].

\bibitem{Braun:1998fs}
   M.~A.~Braun,
   hep-ph/9805394.

\bibitem{Ewerz:2003xi}

   C.~Ewerz,
   hep-ph/0306137.

\bibitem{Ewerz:2005rg}
   C.~Ewerz,
   hep-ph/0511196.

\bibitem{Kwiecinski:1980wb}

   J.~Kwiecinski and M.~Praszalowicz,
   Phys.\ Lett.\  {\bf 94B}, 413 (1980).

\bibitem{Bartels:1980pe}

   J.~Bartels,
   Nucl.\ Phys.\ B {\bf 175}, 365 (1980).

\bibitem{Barakhovsky:1991ra}

   V.~V.~Barakhovsky, I.~R.~Zhitnitsky and A.~N.~Shelkovenko,
   Phys.\ Lett.\ B {\bf 267}, 532 (1991).

\bibitem{Ryskin:1998kt}

   M.~G.~Ryskin,
   Eur.\ Phys.\ J.\ C {\bf 2}, 339 (1998).

\bibitem{Berger:2000wt}

   E.~R.~Berger, A.~Donnachie, H.~G.~Dosch and O.~Nachtmann,
   Eur.\ Phys.\ J.\ C {\bf 14}, 673 (2000).

\bibitem{Ivanov:2001zc}

   I.~P.~Ivanov, N.~N.~Nikolaev and I.~F.~Ginzburg,
   Sci.\ Cult.\ Ser.\ -Phys.\  {\bf 21}, 728 (2002)
   [hep-ph/0110181].

\bibitem{Ginzburg:2002zd}

   I.~F.~Ginzburg, I.~P.~Ivanov and N.~N.~Nikolaev,
   Eur.\ Phys.\ J.\ direct {\bf 5}, no. 1, 002 (2003)
   [hep-ph/0207345].

\bibitem{Hagler:2002nh}
   P.~Hagler, B.~Pire, L.~Szymanowski and O.~V.~Teryaev,
   Phys.\ Lett.\ B {\bf 535} (2002) 117
    Erratum: [Phys.\ Lett.\ B {\bf 540} (2002) 324]
   [hep-ph/0202231].

\bibitem{Machado:2011vh}
M.~V.~T.~Machado,
   Phys.\ Rev.\ D {\bf 86}, 014029 (2012)
   [arXiv:1112.0271 [hep-ph]].

\bibitem{Bolz:2014mya}
   A.~Bolz, C.~Ewerz, M.~Maniatis, O.~Nachtmann, M.~Sauter and
A.~Schoning,
   JHEP {\bf 1501}, 151 (2015)
   [arXiv:1409.8483 [hep-ph]].

\bibitem{Goncalves:2018pbr}
   V.~P.~Goncalves,
   Eur.\ Phys.\ J.\ C {\bf 79}, no. 5, 408 (2019)
   [arXiv:1811.07622 [hep-ph]].

\bibitem{HKMR} 
  L.~A.~Harland-Lang, V.~A.~Khoze, A.~D.~Martin and M.~G.~Ryskin,
  Phys.\ Rev.\ D {\bf 99}, no. 3, 034011 (2019)
  [arXiv:1811.12705 [hep-ph]].

\bibitem{DL} A. Donnachie and P.V. Landshoff, Nucl. Phys. {\bf B231}, 189 (1984); Phys. Lett. {\bf B296}, 227 (1992).

\bibitem{WA102}
  D.~Barberis {\it et al.} [WA102 Collaboration],
  Phys.\ Lett.\ B {\bf 462} (1999) 462
  [hep-ex/9907055].

\bibitem{SC3}  	
L.A. Harland-Lang, V.A. Khoze and M.G. Ryskin,  Eur.Phys.J. {\bf C79} (2019)  39 [arXiv:1810.06567].

\bibitem{FSZ}
  L.  Frankfurt,  V.  Guzey,  M.  Strikman  and  M.  Zhalov,  Phys.  Lett.  {\bf B752} (2016)  51, [arXiv:1506.07150];\\
   C. Ciofi degli Atti, B. Z. Kopeliovich, C. B. Mezzetti, I. K. Potashnikova and I. Schmidt, Phys. Rev. {\bf C84} (2011) 025205 [arXiv:1105.1080].

\bibitem{Woods} R.D. Woods and D.S. Saxon, Phys. Rev. {\bf 95} (1954) 577.

\bibitem{Tarbert} C.M. Tarbert et al., Phys. Rev. Lett. {\bf 112} (2014) 242582.

\bibitem{Jones} A.B. Jones and B.A. Brown, Phys. Rev. {\bf C98} (2014) 067384.

\bibitem{HERA}
  J.~A.~Crittenden,
``Exclusive production of neutral vector mesons at the electron - proton collider HERA,''
  Berlin, Germany: Springer (1997) 100 p
  [hep-ex/9704009].

\bibitem{VDM}
  J.~J.~Sakurai,
  Annals Phys.\  {\bf 11}, 1 (1960);\\
  T.~H.~Bauer, R.~D.~Spital, D.~R.~Yennie and F.~M.~Pipkin,
  Rev.\ Mod.\ Phys.\  {\bf 50}, 261 (1978),
  Erratum: [Rev.\ Mod.\ Phys.\  {\bf 51}, 407 (1979)].

\bibitem{BK} G. Bertsch {\it et al.}, Phys. Rev. Lett. {\bf 47}, 297 (1981);\\
B.Z. Kopeliovich, L.I. Lapidus and A.B. Zamolodchikov, JETP Lett. {\bf 33}, 595 (1981). 

\bibitem{Khoze:2001xm} 
  V.~A.~Khoze, A.~D.~Martin and M.~G.~Ryskin,
  Eur.\ Phys.\ J.\ C {\bf 23}, 311 (2002)
  [hep-ph/0111078].
 
 \bibitem{Kaidalov:2003fw} 
  A.~B.~Kaidalov, V.~A.~Khoze, A.~D.~Martin and M.~G.~Ryskin,
  Eur.\ Phys.\ J.\ C {\bf 31}, 387 (2003)
  [hep-ph/0307064].
  
  
 \bibitem{Harland-Lang:2014lxa} 
  L.~A.~Harland-Lang, V.~A.~Khoze, M.~G.~Ryskin and W.~J.~Stirling,
  Int.\ J.\ Mod.\ Phys.\ A {\bf 29}, 1430031 (2014)
  [arXiv:1405.0018 [hep-ph]].
 
  \bibitem{Khoze:2000jm} 
  V.~A.~Khoze, A.~D.~Martin and M.~G.~Ryskin,
  Eur.\ Phys.\ J.\ C {\bf 19}, 477 (2001)
  Erratum: [Eur.\ Phys.\ J.\ C {\bf 20}, 599 (2001)]
  [hep-ph/0011393].

\bibitem{Bergstrom:1982qv} 
  L.~Bergstrom, G.~Hulth and H.~Snellman,
  Z.\ Phys.\ C {\bf 16}, 263 (1983).

\bibitem{Li:1990sx} 
  Z.~P.~Li, F.~E.~Close and T.~Barnes,
  Phys.\ Rev.\ D {\bf 43}, 2161 (1991).
  
\bibitem{Uehara:2008ep}
  S.~Uehara {\it et al.} [Belle Collaboration],
  Phys.\ Rev.\ D {\bf 78} (2008) 052004
  [arXiv:0805.3387 [hep-ex]].

\bibitem{Breakstone:1986xd} 
  A.~Breakstone {\it et al.} [Ames-Bologna-CERN-Dortmund-Heidelberg-Warsaw Collaboration],
  Z.\ Phys.\ C {\bf 31}, 185 (1986).

\bibitem{Breakstone:1988bm}
  A.~Breakstone {\it et al.} [ABCDHW Collaboration],
  Z.\ Phys.\ C {\bf 40} (1988) 41.

\bibitem{Akesson:1983jz} 
  T.~Akesson {\it et al.} [Axial Field Spectrometer Collaboration],
  Phys.\ Lett.\  {\bf 133B}, 268 (1983).
  
\bibitem{Akesson:1985rn} 
  T.~Akesson {\it et al.} [Axial Field Spectrometer Collaboration],
  Nucl.\ Phys.\ B {\bf 264}, 154 (1986).  
  

\bibitem{Gutierrez:2014yqa}
  G.~Gutierrez and M.~A.~Reyes,
  Int.\ J.\ Mod.\ Phys.\ A {\bf 29} (2014) no.28,  1446008
  [arXiv:1409.8243 [hep-ex]].
  
  \bibitem{Mike} M.~Albrow, invited talk at the  the Low-x workshop, Nicosia, Cyprus, 
26-31 August 2019.

\bibitem {LHCbR} R.~McNulty, [LHCb collaboration], arXiv:1711.06668.

\bibitem{flux}
  G.~Baur and L.~G.~Ferreira Filho,
  Nucl.\ Phys.\ A {\bf 518}, 786 (1990);\\
  K.~Hencken, D.~Trautmann and G.~Baur,
 Z.\ Phys.\ C {\bf 68}, 473 (1995)
  [nucl-th/9503004].
\bibitem{Odd} 
  J.~Bartels, L.~N.~Lipatov and G.~P.~Vacca,
  Phys.\ Lett.\ B {\bf 477}, 178 (2000).

\bibitem{wing}P. Newman and M. Wing, Rev. Mod. Phys. 86,  1037 (2014);
   [arXiv:1308.3368]

\bibitem{HERAf2}T.~Berndt, [H1 collaboration], Acta Phys.\ Polon.\  {\bf B33}, 3499 (2002).

\bibitem{CMS}
  V.~Khachatryan {\it et al.} [CMS Collaboration],
  arXiv:1706.08310 [hep-ex].

\end{document}